# *Controlling friction in a manganite surface by resistive switching*

H. Schmidt,[1] J.-O. Krisponeit,[2,3] K. Samwer,[2] and C. A. Volkert[1,a]

[1] *Institute of Materials Physics, Georg August University, 37077 Göttingen, Germany*

[2] *1st Physics Institute, Georg August University, 37077 Göttingen, Germany*

[3] *Institute of Solid State Physics, University of Bremen, 28359 Bremen, Germany*

We report a significant change in friction of a $La_{0.55}Ca_{0.45}MnO_3$ thin film measured as a function of the materials resistive state under ultrahigh vacuum conditions at room temperature by friction force microscopy. While friction is high in the insulating state, it clearly changes to lower values if the probed local region is switched to the conducting state via nanoscale resistance switching. Thus we demonstrate active control of friction without having to change the temperature or pressure. Upon switching back to an insulating state the friction increases again. The results are discussed in the framework of electronic friction effects and electrostatic interactions.

Friction is a complex phenomenon that occurs between two bodies at a sliding contact. Despite the fact that it often can be described by straight-forward empirical relations its fundamental cause is by no means simple. With the advent of the atomic force microscope (AFM), understanding and controlling nanoscale friction has become one of the major interests in modern tribology. Several literature studies [1–5] report a clear change in measured nanoscale friction when the electronic state of the material is altered. Increases are observed in the non-contact dissipation of Nb [4] and the contact friction of YBCO [5] as is the materials are heated through their superconducting transitions. Contact friction measurements on Si [3] and GaAs [1] semiconductors demonstrated a strong dependence on the charge carrier density. Only recently Kim et al. demonstrated that the contact friction of $VO_2$ is strongly increased on heating through the metal-to-insulator transition (MIT) to the metallic state [2]. However, most of these studies use temperature and/or contact pressure to control the materials electronic state or need to permanently apply a bias voltage between the tip and the sample to observe changes in friction. Here, we report an active control of friction where there is no need to control the temperature, contact pressure or bias voltage during the friction experiment. Perovskite manganites offer an interesting material to investigate and control nanoscale friction by changing their electrical resistance. Their resistance behavior shows a variety of interesting phenomena, including the well-known temperature driven metal-to-insulator transition [6] and the colossal magnetoresistance (CMR) effect [7] as well as bipolar nanoscale resistive switching driven by electric fields [8]. Here we use the last to reversibly alter the resistive state of a nanoscale region of a perovskite manganite surface to investigate the effect on friction using AFM-based methods. This gives us the opportunity to bypass temperature-dependent contact mechanics effects that can dominate nanoscale friction at low temperatures [9].

---

[a] Author to whom correspondence should be addressed. Electronic mail: volkert@ump.gwdg.dev

The manganite $La_{0.55}Ca_{0.45}MnO_3$ was deposited as a thin film on a MgO substrate with (100)-orientation by the metalorganic-aerosol deposition technique [10]. Small angle x-ray scattering gives a film thickness of approximately 44 nm and $\Theta/2\Theta$ x-ray diffraction experiments confirm (100) epitaxial growth on the substrate (see Fig. S1). The film sheet resistance vs. temperature behavior probed using four-point measurements shows a clear metal-to-insulator transition at $T_{MI}$ = 245 K (see Fig. S2). Both the lattice constant (3.87(3) Å) and the transition temperature are consistent with a Ca doping fraction of 0.45 as expected from the deposition conditions. Hence the manganite film is a paramagnetic insulator at room temperature and a ferromagnetic metal below $T_{MI}$ [6]. As shown previously for equivalent samples, the near-surface region of such films can be switched on the nanoscale between two resistive states by conductive atomic force microscopy (C-AFM) [11,12].

Friction experiments and resistive switching have been performed in a vacuum chamber with an Omicron VT-AFM/STM at room temperature and a base pressure of $10^{-10}$ mbar. The cantilever normal and torsional spring constants are 0.2 N/m and 23.49 N/m, respectively [13,14]; lateral forces were calibrated using analytical procedures described in Ref. [15]. The measurements presented in the following combine friction force microscopy (FFM) and conductive atomic force microscopy (C-AFM) using a Pt-coated silicon cantilever. This allows us to simultaneously measure the friction and to switch the resistive state of the manganite back and forth between the metallic (low resistance state, LRS) and insulating states (initial state IS, high resistance state HRS) by applying a voltage larger than the switching voltage $|V_C| \approx 3$ V to the tip. Additionally, the C-AFM gives us the opportunity to resolve the resistive state by recording current maps at lower tip voltages during friction measurements. In order to avoid plastic protrusions that can be generated by Joule heating under application of the large positive switching voltages used for generating the LRS [11], we rapidly scanned the area of interest several times under an applied bias rather than using prolonged voltage pulses at a fixed cantilever position. Using this method, we avoided any noticeable changes in the surface morphology in our experiments. Nor was a correlation between rms surface roughness (typically around 1 nm) and friction or current observed.

In each experiment, the friction forces were measured in a 500x500 nm$^2$ region on the sample surface by performing 600 friction loops, i.e. by recording the lateral forces experienced by the cantilever tip while scanning forward and backward traces over the sample surface. At the same time, the surface topography was recorded from the vertical deflection of the cantilever. The current was also recorded during the friction loops by applying a voltage of 0.1 V which is well below the threshold needed for switching to probe the materials resistive state. First, friction, topography, and current maps were obtained from a 500x500 nm$^2$ region of the specimen in the insulating initial state (IS). Then the near surface region was switched to the conducting LRS by scanning over the same region with a tip voltage of 3.5 V which is sufficiently high to alter the materials resistive state. After obtaining friction, topography, and current maps from the switched state, the region was switched back to an insulating HRS using a tip voltage of -3.5 V, and friction, topography, and current maps were obtained once again. The normal force load was kept constant at $F_N = 5$ nN and the scan velocity was 2000 nm/s. This experiment was repeated for different local regions of interest on the specimen.

**FIGURE 1**

Fig. 1(a-c) show topography and lateral force maps of the three resistive states (IS, HRS, LRS) for a representative region on the sample. Due to some drift between the tip and specimen during the resistive switching procedure, the scanned regions are slightly displaced relative to each other and have been aligned using the topography maps. The common height and force scales for the three sets of maps show clearly that there are no significant changes in topography, but that the lateral forces are strongly reduced in the conducting LRS (Fig. 1(b)). Large variations in the lateral forces are observed within each map that are on a similar length scale as the topography (Fig. 1(a-c)). We used the current maps recorded simultaneously at a bias voltage V = 0.1 V to ensure that the resistive state was induced everywhere in the scanned area However, because the current amplifier reached its saturation limit of 50 nA even at a low bias (0.1 V) in the LRS and because the currents in the insulating states (IS and HRS) are below the detection limit, we obtain only threshold information.

**FIGURE 2**

The dramatic change in friction with resistive state is illustrated most clearly by the friction loops. Fig. 2 shows average friction loops for each resistive state, which have been calculated by averaging the lateral forces for all 600 forward and backward traces within a single region of interest. The x-shift of around 50 nm of the LRS loop relative to the IS and HRS is due to drift that occurred during resistive switching. A clear hysteresis in the lateral force and thus large friction is observed for the initial and switched insulating states (IS and HRS), while the conducting state (LRS) shows almost no hysteresis and thus low friction.

Large variations in lateral force are observed along the forward and backward traces that are correlated with sample position (Fig. 2). Furthermore, these local variations are mostly unchanged on switching to the conducting state and on switching back to the insulating state. The most likely cause for these fixed variations in friction are gradients in the local sample surface height dh/dx which contribute, additively with the actual friction force, to the measured lateral force [16]. This was confirmed by the fact that surface height gradient maps overlaid almost exactly on the lateral force maps. It also justifies the method of calculating friction forces from lateral force maps, namely by first x-shifting the forward and backward traces relative to each other to account for the rotation of the tip on reversing the scan direction (around 15 nm in Fig. 2), and then by subtracting the forces obtained from the forward and backward traces of a friction loop, thereby minimizing contributions from topography.

**FIGURE 3**

In order to confirm the robustness of the effect of the resistive state on friction, the above experiment was repeated for a number of different regions of interest on the specimen (Figure 3). The average friction force for each region of interest and resistive state was calculated by shifting and subtracting the forward and backward lateral forces and dividing by two. Despite some scatter, the friction for all regions is clearly reduced on switching from the insulating IS to the conducting LRS and increases on switching back to the HRS. However, it is found that that the friction force of the IS is not fully recovered on switching back from the

LRS to the HRS. This observation is consistent with previous studies [11] where resistance fatigue was observed when switching a manganite thin film back and forth between its resistive states, particularly on the first switching cycle. Specifically, the difference between the resistances of the two states decreased with switching cycle number. Furthermore, Krisponeit et al. reported local variations in the threshold voltage and the magnitude of the resistance change on switching [11] as well as in the time stability of the LRS [12]. Therefore, the variations from region to region in the initial values of the friction and in the magnitude of the friction change on switching, likely have the same origins as the local resistance variations. Whether the common cause of both resistance and friction variations are local variations in composition, local variations in the thickness of the oxygen deficient surface layer (see below), or some other effect, is not yet clear. In any case, a clear correlation between resistance state and friction has been demonstrated.

According to theory (see for example [17]), sliding friction over a metallic substrate should cause electronic excitations, giving rise to dissipation in excess of that due to phonons. Estimates of the magnitude of this effect yield values no more than $F/v = 10^{-9}$ Ns/m, where F is the excess friction force and v the scanning velocity [1]. Most literature studies do show an excess friction as a result of adding charge carriers [1,3] or due to heating from the superconducting to normal state [4,5] or from an insulator to metal state [2]. However, the magnitude of the experimental effect for contact friction measurements is many of orders of magnitude larger than predicted by theory. Even if one considers stick-slip motion of the tip, which would lead to substantially higher sliding speeds [18], theoretical predictions of electronic effects cannot explain the observed excess friction. Only in cases where the phononic friction is sufficiently low, such as for non-contact dissipation measurements [4] and dissipation during adsorbate sliding [19,20] is there at least qualitative agreement between theory and experiment.

The disagreement between theory and experiment has largely been reconciled by considering contact electrification and electrostatic effects. Particularly in oxides and other insulating materials, it has often been proposed that electrostatic forces from trapped charges may overwhelm other effects [1–3,5,21]. For example, Qi et al. [1] were able to obtain quantitative agreement with their experimental excess friction, which was on the order of $10^{-5}$ Ns/m, by accounting for the electrostatic forces from trapped charges in the surface oxide layer. Altfeder et al. [5] found even larger excess dissipation levels in the range of $10^{-2}$–$10^{-3}$ Ns/m between the normal and superconducting phase of YBCO, and attributed it to contact electrification and electrostatic effects in the oxygen depleted surface layer of the normal phase. Recently, Kim et al. [2] also employed this argument to explain the large excess friction observed on transforming $VO_2$ from the insulating to metallic phase by either increasing the temperature or increasing the stress. Specifically, they attributed the excess friction to trapped charges in the surface $V_2O_5$ dielectric layer on the metallic $VO_2$ domains, leading to higher Coulomb attraction between the tip and sample.

In our case, we observe a *reduction* in friction on the order of $10^{-3} - 10^{-2}$ Ns/m when *switching* the manganite thin film from the insulating to the conducting state. This is comparable to the magnitude of friction reduction observed in other systems and the *excess friction associated with the insulating state* is perhaps not so difficult to explain. A similarly large *increase* is observed in other systems on transforming from the insulating to the conducting state [2] and is explained by charges trapped in an oxide layer on the metallic phase. Despite the fact that our effect goes in the opposite direction it can be explained by the same phenomenon. In contrast to the materials in the previous studies, the conducting LRS of manganite does not form a surface oxide where charges can be trapped. Instead, as-deposited doped manganites often have an oxygen deficient surface region which leads to charge transfer from the bulk to the surface layer, contributing to surface-dipole formation and the formation of $Mn^{3+}$ ions at the surface [22,23]. Furthermore surface dipole formation is also observed in literature for most alkyl halide [24]. Therefore, we propose that the high friction in the electrically insulating IS and HRS may result from strong electrostatic interactions between the tip and the $Mn^{3+}$ ions at the surface of the specimen. When we switch to the LRS, a structural transition leads to increased electron/hole mobility by changing the Mn-O-Mn bond angle [11]. No surface dipoles should be present in this state which decreases the electrostatic part of the tip-sample interaction significantly. It may be that the larger variation in friction of the insulating state relative to the metallic state (Fig. 3) results from local variations in the distribution and density of Mn or O ions. Thus an electrostatic origin for the observed excess friction in the insulating state of the manganite film is not only plausible, it provides an explanation that is consistent with the previous literature studies, where friction is increased due to electrostatic forces acting between the tip and charges that are created or accumulated at the materials surface [1–3,5].

In conclusion, we report have clearly demonstrated control of nanoscale friction by resistive switching the local resistive state of the manganite without having to vary the temperature or pressure. A reproducible and reversible change in friction is observed when the manganite film is resistively switched between insulating and conducting states. The excess friction of the insulating state is explained by electrostatic effects due to trapped charges in a surface oxygen deficient layer, in close analogy with the literature reports of excess friction measured in metals with surface oxide layers.

This work is part of the CRC 1073 (project A01) funded by the Deutsche Forschungsgemeinschaft (DFG). JOK acknowledges support by the Institutional Strategy of the University of Bremen, funded by the German Excellence Initiative. The authors thank A. Belenchuk for sample preparation and C. Meyer for x-ray diffraction characterization, and gratefully acknowledge helpful discussions with B. Damaschke, J. Krim, and V. Moshnyaga.

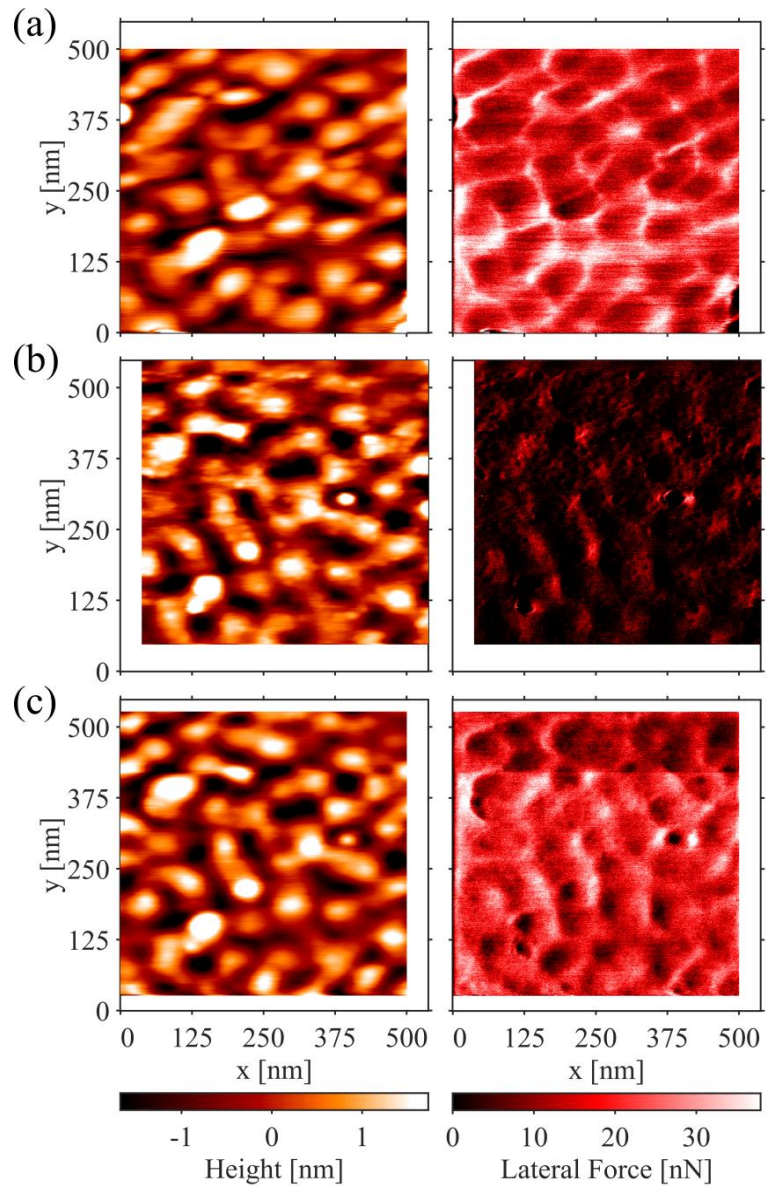

**FIGURE 1:** AFM-based measurements of $La_{0.55}Ca_{0.45}MnO_3$. (a-c) Topography (left) and lateral force (right) maps obtained from forward traces measured on (a) the initial insulating state (IS), (b) after resistively switching to the metallic state (LRS), and (c) after resistively switching back to the insulating state (HRS). During the scans current maps were recorded at a bias voltage V = 0.1 V obtaining threshold information about the resistive state. The sample was either fully insulating (IS, HRS) or fully conductive (LRS) with currents below the detection limit of 0.1 nA or above the detection limit of 50 nA, respectively.

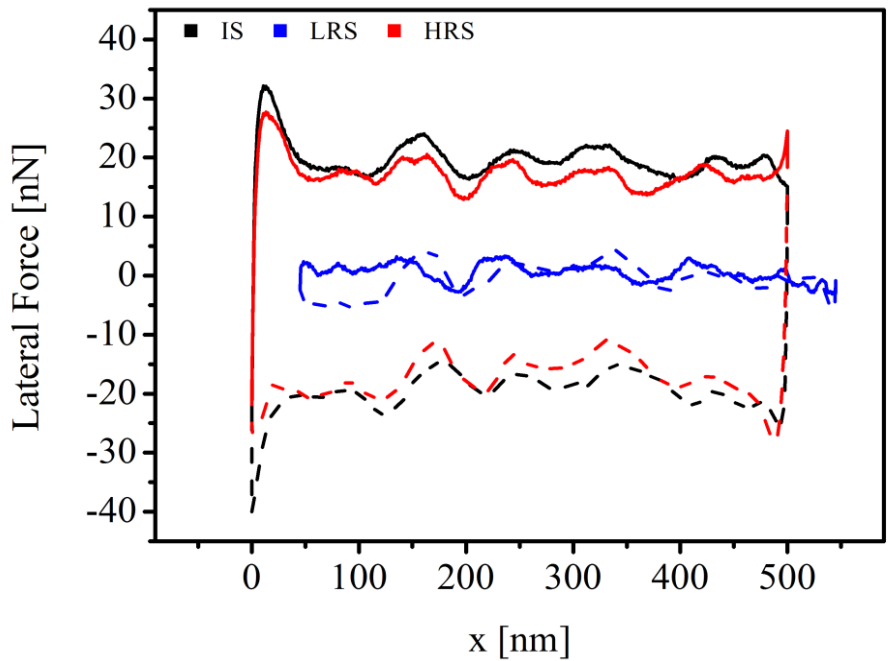

**FIGURE 2:** (color online) Average friction loops for the insulating states (IS = black; HRS = red) and the conducting state (LRS =blue). The friction is represented by the difference between the forward (solid line) and backward (dashed line) trace lateral forces divided by two. Thus, the friction is high for the insulating states and low for the conducting state. The slight offset (ca. 15 nm) in the x-direction between the forward and backward scans is due to rotation of the tip on reversing the scan direction.

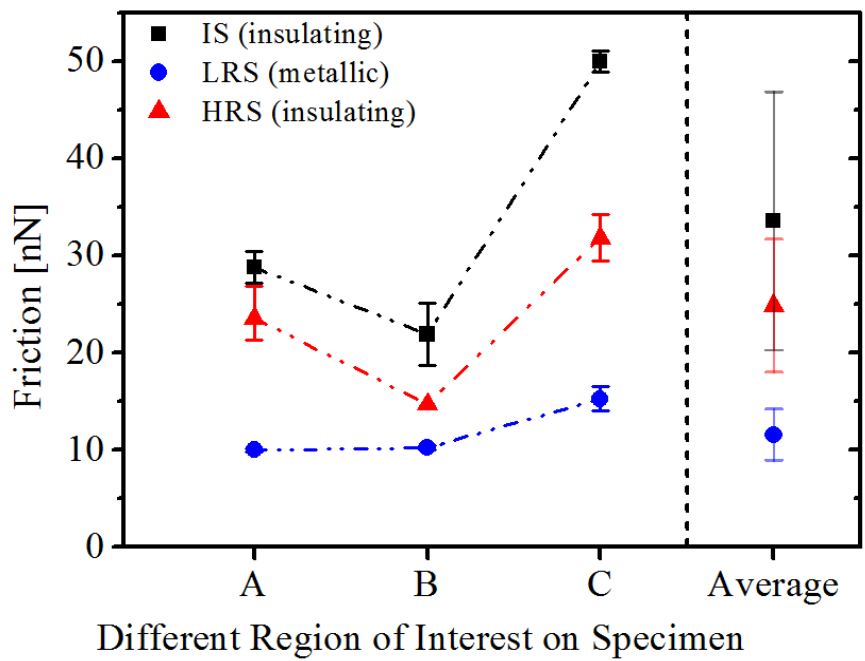

**FIGURE 3:** (color online) Friction of the insulating state is consistently higher than of the metallic state in three different regions of the $La_{0.55}Ca_{0.45}MnO_3$ film. The left hand portion of the plot shows average values (symbols) and ranges (bars) for friction of each resistive state in the three regions. The means and standard deviation of the position averaged values of the friction for each resistive state are shown in the right hand portion of the plot.

## *Supporting Information*

for *Controlling friction in a manganite surface by resistive switching*

by H. Schmidt, J.-O. Krisponeit, K. Samwer, and C. A. Volkert

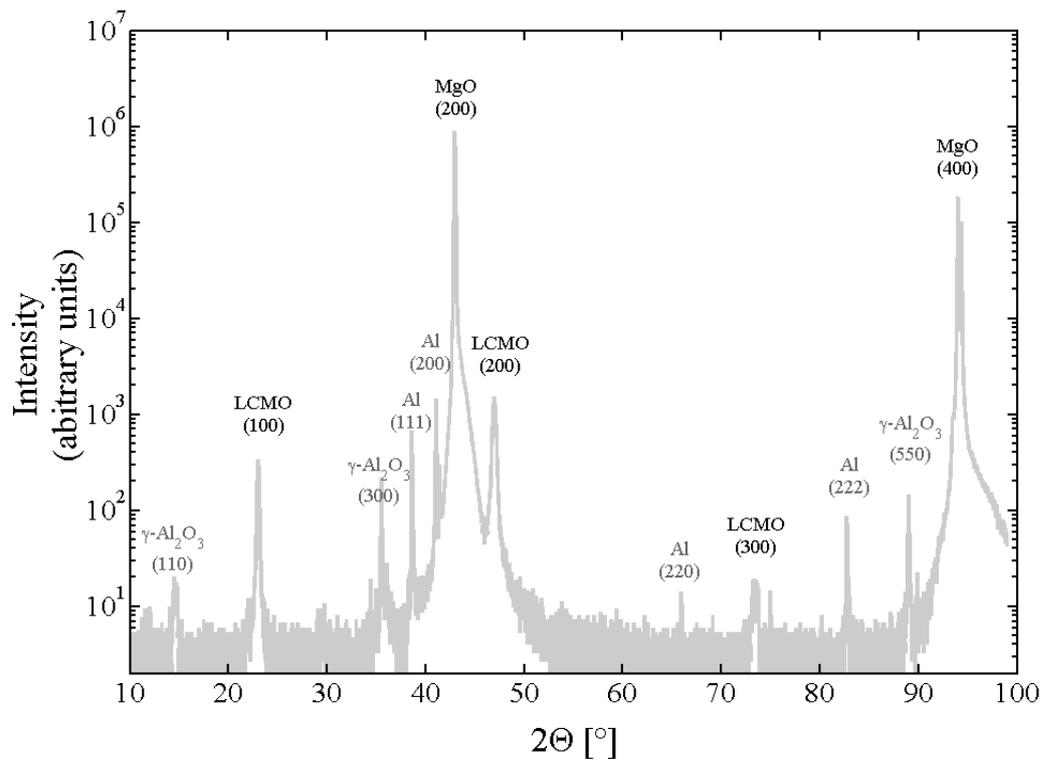

**FIGURE S1:** X-ray diffraction experiments confirm (100) epitaxial growth of the La$_{0.55}$Ca$_{0.45}$MnO$_3$ thin film on the MgO substrate. The Al and Al$_2$O$_3$ peaks belong to the sample mounting.

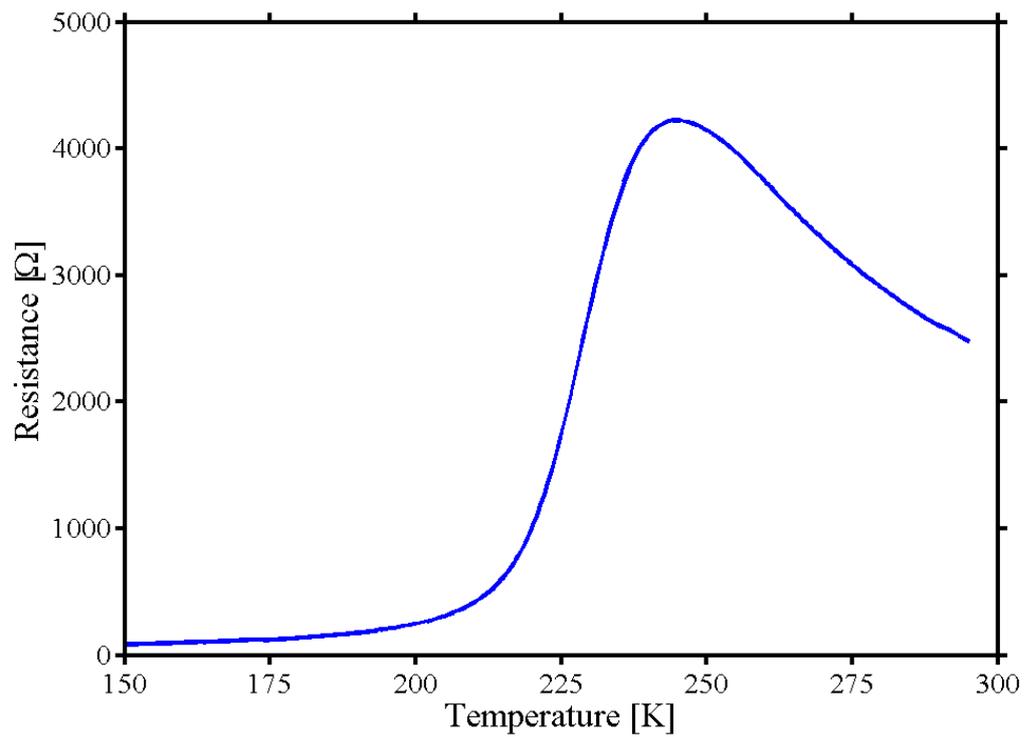

**FIGURE S2:** Resistance as a function of the $La_{0.55}Ca_{0.45}MnO_3$ thin films temperature from a four-point measurement performed on a Quantum Design Physical Property Measurement System shows a clear metal-to-insulator transition at $T_{MI} = 245$ K.